\documentclass[12pt]{article}
\usepackage{amssymb}
\usepackage{amsmath}
\usepackage[space]{cite}
\usepackage{algorithm}
\usepackage[noend]{algpseudocode}
\usepackage[normalem]{ulem}
\usepackage{url}
\usepackage{graphicx}
\usepackage{adjustbox}
\usepackage{caption}
\usepackage{subcaption}

\usepackage{quantikz}
\usetikzlibrary{circuits.logic.US}

\newcommand{\be}{\begin{eqnarray}}
\newcommand{\ee}{\end{eqnarray}}
\newcommand{\non}{\nonumber}
\newcommand{\id}{\mathbb{I}}

\usepackage{quantikz}
\usetikzlibrary{circuits.logic.US}
\tikzset{
    gateO/.style={
        draw,
        circle,
        minimum width=0.5em,
        inner sep=2pt    }
}
\DeclareExpandableDocumentCommand{\gateO}{O{}{m}}{|[gateO,#1]| {#2} \qw}

\begin{document}

\begin{titlepage}
\vspace{.5in}
\begin{center}

{\LARGE Quantum counting, and a relevant sign}\\
\vspace{1in}
\large Natalie Chung\footnote{Doral Academy Preparatory High School,
11100 NW 27th St, Doral, FL 33172 USA {\tt nataliechung05@gmail.com}} 
and Rafael I. Nepomechie\footnote{
Department of Physics, P.O. Box 248046, University of Miami,  
Coral Gables, FL 33124 USA  {\tt  
nepomechie@miami.edu}}
\end{center}

\vspace{.5in}

\begin{abstract}
Two indispensable algorithms in an introductory course on Quantum 
Computing are Grover's search algorithm and quantum phase estimation. 
Quantum counting is a simple yet beautiful blend of these two 
algorithms, and it is therefore an attractive topic for a student 
project in such a course. However, a sign that is irrelevant when 
implementing Grover's algorithm becomes relevant. We briefly review 
these algorithms, highlighting the aforementioned sign.
\end{abstract}

\end{titlepage}

\setcounter{footnote}{0}

\section{Introduction}\label{sec:intro}

Two indispensable algorithms in an introductory undergraduate course on Quantum
Computing are Grover's search algorithm \cite{Grover:1996rk, Grover:1997fa} and quantum phase estimation
(QPE) \cite{Kitaev:1995qy}.  Quantum counting \cite{Boyer:1996zf, Brassard:1998vj} is a
simple yet beautiful blend of these two algorithms, and it is
therefore an attractive topic for a student project in such a course.
However, a sign that is irrelevant when implementing Grover's
algorithm becomes relevant.

We start by briefly reviewing in Sec. \ref{sec:Grover} Grover's algorithm for a single 
marked element, and its extension \cite{Boyer:1996zf} to the case of multiple marked 
elements. QPE is briefly reviewed in Sec. \ref{sec:QPE}. The key part of this paper
is Sec. \ref{sec:QC}, where we review quantum counting and highlight the 
aforementioned sign. We finish in Sec. \ref{sec:discuss} with a brief conclusion.

\section{Grover's algorithm}\label{sec:Grover}

Grover's search algorithm, one of the ``crown jewels'' of Quantum
Computing, has been widely described in detail, see e.g. \cite{Mermin:2007}.  We content ourselves here with reminding the reader of the main steps, which also serves to set out our notations
and conventions. We first treat in Sec. \ref{sec:onemarked} the familiar case of a single marked element, and then consider in Sec. \ref{sec:multiplemarked} the perhaps less-familiar case of multiple marked elements.

\subsection{One marked element}\label{sec:onemarked}

Let $x$ and $a$ be $n$-bit integers, and $f(x)$ the search function
\begin{equation}
	f(x) = \begin{cases}
	0 & \text{if }\quad x \ne a \\
	1 & \text{if }\quad x = a
	\end{cases}
\end{equation}
The ``marked element'' $a$ is the object of our search. We define the 
corresponding unitary operator $U_{f}$, the so-called standard 
protocol acting on the $n$-qubit ``input register'' $|x\rangle$ and 
the 1-qubit  ``output register'' 
$|y\rangle$, by
\begin{equation}
	U_{f}\left(|y\rangle |x\rangle \right) = |y \oplus f(x) \rangle 
	|x\rangle \,.
	\label{Ufunction}
\end{equation}
The initial state of the output register is taken to be
\begin{equation}
	|y\rangle = H |1\rangle = \frac{1}{\sqrt{2}}\left( |0\rangle 
	-|1\rangle\right) \,,
\end{equation}
and therefore
\begin{equation}
	U_{f}\left(H|1\rangle |x\rangle \right) = (-1)^{f(x)} H|1\rangle 
	|x \rangle \,.
	\label{Ufaction}
\end{equation}
We define the operator $V$ on the input register by
\begin{equation}
	V |x\rangle = (-1)^{f(x)} |x\rangle 
	= \begin{cases}
	\ \ |x \rangle & \text{if }\quad x \ne a \\
	-|x \rangle & \text{if }\quad x = a
	\end{cases} \,,
\end{equation}
which can be realized as
\begin{equation}
	V = 1 - 2|a\rangle\langle a | \,,
	\label{Vop}
\end{equation}
so that \eqref{Ufaction} can re-expressed as
\begin{equation}
	U_{f}\left(H|1\rangle |x\rangle \right) = H|1\rangle 
	V |x \rangle\,.
	\label{Ufaction2}
\end{equation}
The output register remains in the state $H|1\rangle $ throughout the 
algorithm, and so we henceforth focus only on the input register.

The initial state of the input register is taken to be
\begin{equation}
	|\phi \rangle \equiv H^{\otimes n} |0\rangle^{\otimes n} = 
	\frac{1}{2^{n/2}} \sum_{x=0}^{2^{n}-1} |x\rangle \,.
	\label{phi}
\end{equation}	

\begin{figure}[htb]
 	\centering
 	\includegraphics[width=0.3\hsize]{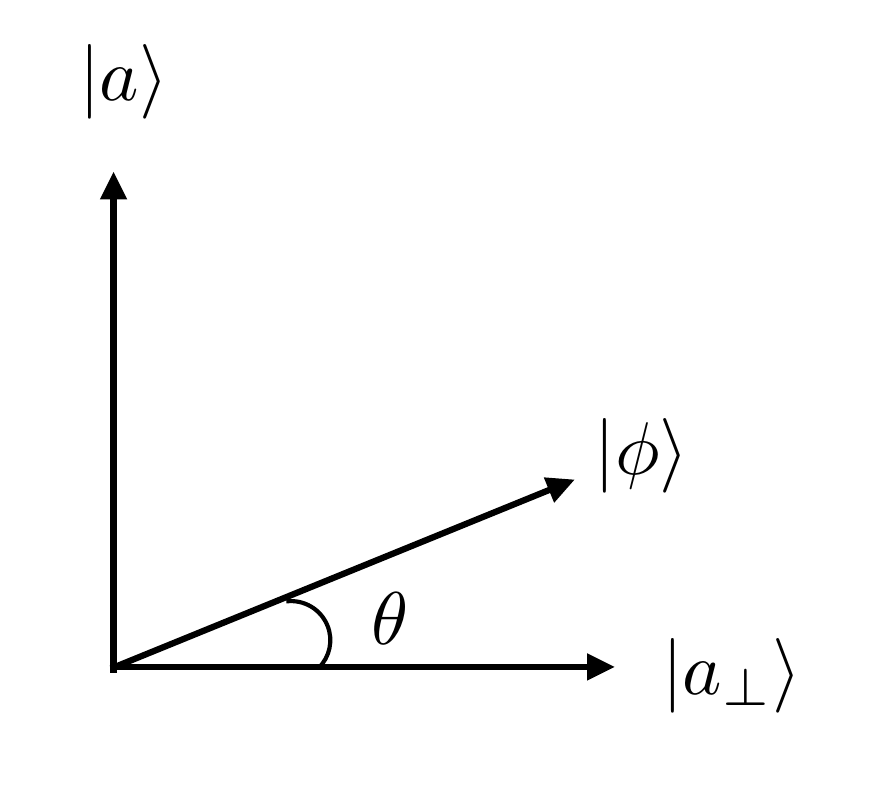}
	\caption{The plane spanned by $|a\rangle$ and $|\phi\rangle$}
	\label{fig:aaperp}
 \end{figure}

The kets $|a \rangle$ and $|\phi \rangle$ determine a (real) plane; we 
define a normalized ket $|a_{\perp} \rangle$ in this plane that is 
perpendicular to $|a \rangle$, i.e. $\langle a|a_{\perp} \rangle = 0$.
We define $\theta$ as the angle between $|a_{\perp} \rangle$ and $|\phi 
\rangle$, as in Fig. \ref{fig:aaperp}. We therefore have
\begin{equation}
	|\phi \rangle = \cos \theta\, |a_{\perp} \rangle + \sin \theta\, |a 
	\rangle  \,.
\label{phiaaperp}
\end{equation}	
We observe from \eqref{phi} that $\langle a | \phi \rangle = 
\frac{1}{2^{n/2}}$; and from \eqref{phiaaperp}
that $\langle a | \phi \rangle  = 
\sin \theta$. Hence,
\begin{equation}
	\sin \theta = \frac{1}{2^{n/2}} \,.
	\label{sintheta}
\end{equation}

We define the unitary operator $W$, the so-called diffuser, by
\begin{equation}
	W = 2|\phi\rangle\langle \phi | -1  \,.
	\label{Wop}
\end{equation}
By explicit computation using \eqref{Vop}, \eqref{phiaaperp} and 
\eqref{Wop}, we obtain
\begin{align}
	\left(W V\right) |a_{\perp} \rangle &= \cos(2 \theta)\, |a_{\perp} \rangle 
	+ \sin(2\theta)\, |a \rangle  \,, \non \\
	\left(W V\right) |a \rangle &= -\sin(2 \theta)\, |a_{\perp} \rangle 
	+ \cos(2\theta)\, |a \rangle  \,.
\label{WV}
\end{align}	
We see that $WV$ can be represented in the basis $|a_{\perp} 
\rangle\,, |a \rangle $ by the matrix
\begin{equation}
WV = \begin{pmatrix}
\cos(2\theta) & -\sin(2\theta) \\
\sin(2\theta) & \cos(2\theta) 
\end{pmatrix} = e^{- i 2 \theta\, Y} \,,
\label{WVmat}
\end{equation}
which rotates any ket in the plane by the  angle $2\theta$ 
counterclockwise (i.e, from $|a_{\perp} 
\rangle$ to $|a \rangle $). After $k$ such Grover rotations,
the input register is in the state
\begin{equation}
	\left(WV \right)^{k}\, |\phi\rangle = 
	\cos\left((2k+1)\theta\right)\, |a_{\perp} \rangle 
	+\sin\left((2k+1)\theta\right)\, |a\rangle \,.
\label{WVk}	
\end{equation}
The number $k$ is chosen such that the result \eqref{WVk} is the 
sought-after ket $|a\rangle$;  measuring the input register will then give $a$ with 
probability 1. To this end, we set $(2k+1)\theta = \pi/2$, and thus 
$k=\frac{\pi}{4\theta} - \frac{1}{2}$. We see from 
\eqref{sintheta} that $\theta \approx \sin \theta = 
\frac{1}{2^{n/2}}$ for large $n$. Hence, the required number of Grover rotations is
\begin{equation}
	k \approx \frac{\pi}{4} \sqrt{2^{n}} \,.
	\label{k1}
\end{equation}

It is not difficult to show that the $W$ operator \eqref{Wop} can be 
re-expressed as
\begin{equation}
	W = - H^{\otimes n} X^{\otimes n} \left(c^{n-1} Z \right) 
	X^{\otimes n} H^{\otimes n} \,,
	\label{Wcircuit}
\end{equation}
where $c^{n-1} Z$ denotes the $(n-1)$-fold controlled $Z$ gate. 
This form for $W$ is useful for implementing Grover's algorithm on, 
say,  a quantum simulator. For the purpose of determining $a$, the 
minus sign in \eqref{Wcircuit} is irrelevant, so it is 
typically dropped. (Note that multiplication by $-1$ is
not a standard gate. Nevertheless, this sign could be implemented using a product of standard 1-qubit gates, for example $Z X Z X = -\id$.) That is, when simulating Grover's algorithm,
typically one uses 
\begin{equation}
    \tilde{W} \equiv - W = H^{\otimes n} X^{\otimes n} \left(c^{n-1} Z \right) 
	X^{\otimes n} H^{\otimes n} 
 \label{Wtilde}
\end{equation}
instead of \eqref{Wcircuit}.
We will revisit this sign in Sec. \ref{sec:QC}.

\subsection{Multiple marked elements}\label{sec:multiplemarked}

Suppose we want to search now for a set $S=\{ a_{1}, a_{2}, \ldots, a_{m}\}$ of 
$m$ marked elements. The standard protocol $U_{f}$ is still given by 
\eqref{Ufunction}, except that the search function is now given by
\begin{equation}
	f(x) = \begin{cases}
	0 & \text{if }\quad x \notin S \\
	1 & \text{if }\quad x \in S
	\end{cases} \,.
\end{equation}
The operator $V$ is therefore now defined by
\begin{equation}
	V |x\rangle = (-1)^{f(x)} |x\rangle 
	= \begin{cases}
	\ \ |x \rangle & \text{if }\quad x \notin S \\
	-|x \rangle & \text{if }\quad x \in S
	\end{cases} \,,
\end{equation}
and is given by
\begin{equation}
	V = 1 - 2\sum_{j=1}^{m}|a_{j}\rangle\langle a_{j} | \,.
	\label{Vop2}
\end{equation}

\begin{figure}[htb]
 	\centering
 	\includegraphics[width=0.3\hsize]{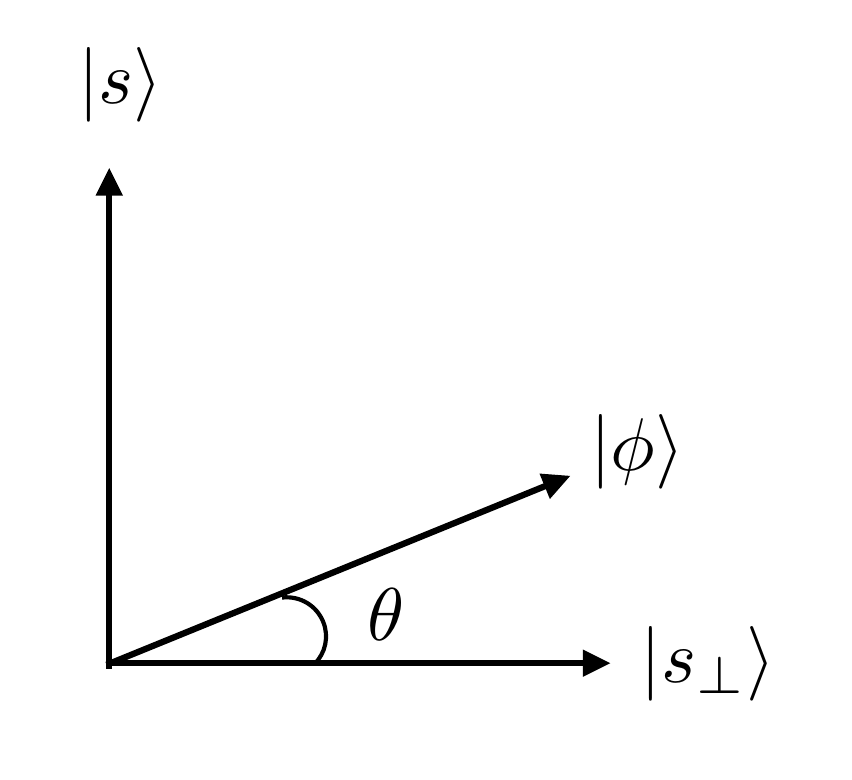}
	\caption{The plane spanned by $|s\rangle$ and $|\phi\rangle$}
	\label{fig:ssperp}
 \end{figure}

The key insight is to consider the plane spanned by 
$|s \rangle$ and $|\phi \rangle$, where $|s \rangle$ is
an equal-weight superposition of all the $|a_{j}\rangle$'s
\begin{equation}
	|s\rangle = \frac{1}{\sqrt{m}}\sum_{j=1}^{m}|a_{j}\rangle \,.
	\label{sket}
\end{equation}
We define a vector $|s_{\perp} \rangle$ in this plane that is 
perpendicular to $|s \rangle$, i.e. $\langle s|s_{\perp} \rangle = 0$;
moreover, we define $\theta$ as the angle between $|s_{\perp} \rangle$ and $|\phi 
\rangle$, see Fig. \ref{fig:ssperp}. We therefore have
\begin{equation}
	|\phi \rangle = \cos \theta\, |s_{\perp} \rangle + \sin \theta\, 
	|s \rangle  \,.
\label{phissperp}
\end{equation}	
We observe from \eqref{sket} and \eqref{phi} that $\langle s | \phi \rangle = 
\sqrt{\frac{m}{2^n}}$; and from \eqref{phissperp}
that $\langle s | \phi \rangle  = 
\sin \theta$. Hence,
\begin{equation}
	\sin \theta = \sqrt{\frac{m}{2^n}} \,.
	\label{sintheta2}
\end{equation}

It is easy to check that $V |s\rangle = - |s\rangle$; furthermore, 
$\langle a_{j} | s_{\perp}\rangle = 0$, and hence 
$V |s_{\perp}\rangle =  |s_{\perp}\rangle$. We see that, in the
plane of $|s_{\perp} \rangle$ and $|s \rangle$, $V$ acts as 
\begin{equation}
	V = 1 - 2|s\rangle\langle s | \,,
	\label{Vop3}
\end{equation}
cf. \eqref{Vop}.
The computation of $WV$ on $|s_{\perp} \rangle$ and $|s \rangle$ is 
therefore the same as the previous computation of $WV$ on $|a_{\perp} \rangle$ 
and $|a \rangle$, respectively \eqref{WV}; and in the basis $|s_{\perp} 
\rangle\,, |s \rangle$, $WV$ is given by the same matrix \eqref{WVmat}.
After $k$ Grover rotations,
the input register is in the state
\begin{equation}
	\left(WV \right)^{k}\, |\phi\rangle = 
	\cos\left((2k+1)\theta\right)\, |s_{\perp} \rangle 
	+\sin\left((2k+1)\theta\right)\, |s\rangle \,.
\label{WVk2}	
\end{equation}
Choosing again $(2k+1)\theta = \pi/2$, so that
\begin{equation}
	k \approx \frac{\pi}{4} \sqrt{\frac{2^{n}}{m}} \,,
	\label{k2}
\end{equation}
the result \eqref{WVk2} is 
the ket $|s\rangle$ \eqref{sket};  measuring the input register will 
then give any one of the $a_{j} \in S$ with equal probability.
Of course, \eqref{k2} reduces to \eqref{k1} for $m=1$.

In order to use this algorithm, we must know the number of marked
elements ($m$), which enters into the formula \eqref{k2} for the
number of Grover rotations.  What do we do if we do {\it not} know $m$ ahead of time?  One
way to proceed, the so-called quantum counting considered in Sec. 
\ref{sec:QC}, involves using QPE.

\section{Quantum phase estimation}\label{sec:QPE}

\begin{figure}[htb]
	\centering
\begin{adjustbox}{width=0.8\textwidth}
\begin{quantikz}
\lstick[wires=5]{$\ket{0}^{\otimes t}$}
& \gate{H} & \ctrl{1}\vqw{5} & \qw & \qw \ldots & \qw & \qw  \slice{1} 
& \gate[wires=5, nwires=3]{U_{FT}^{-1}}  \slice{2} & \qw & \rstick[wires=5]{}  \\
& \gate{H} & \qw & \ctrl{1}\vqw{4} & \qw \ldots & \qw & \qw &  & \qw 
& \\
\vdots & & & &   & & &  & & & |[meter]|\\
& \gate{H} & \qw & \qw & \qw \ldots  & \ctrl{1}\vqw{2} & \qw & & \qw &\\
& \gate{H} & \qw & \qw  & \qw \ldots  & \qw & \ctrl{1}\vqw{1} & & \qw 
& \\
\lstick{$|\psi\rangle$}& \qw 
&\gate{{\mathcal U}^{2^0}} 
&\gate{{\mathcal U}^{2^1}} & \qw \ldots  & \gate{{\mathcal U}^{2^{t-2}}} 
& \gate{{\mathcal U}^{2^{t-1}}} & \qw & \qw &
\end{quantikz}
\end{adjustbox}
\caption{Circuit diagram for QPE}
\label{fig:QPE}
\end{figure}
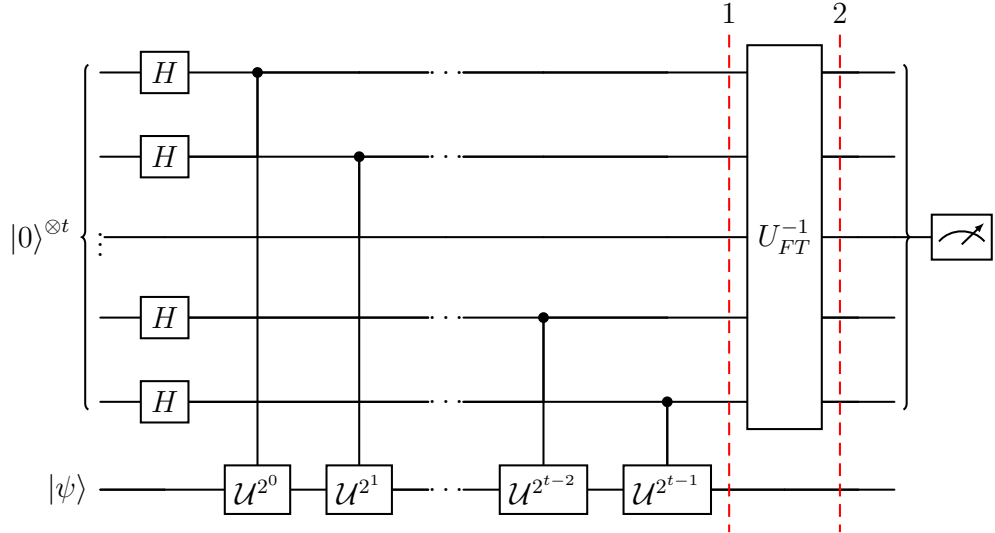

QPE is another important quantum algorithm with many uses, including
an elegant formulation of Shor's celebrated period-finding algorithm 
\cite{Shor:1994jg}.
Given a unitary operator ${\mathcal U}$ and an eigenvector $|\psi\rangle$, 
whose corresponding eigenvalue necessarily has the form 
$e^{i \alpha}$ with $\alpha$ real, 
\begin{equation}
	{\mathcal U}\, |\psi\rangle = e^{i \alpha} |\psi\rangle \,, 
\end{equation}
QPE provides an estimate for $\alpha$. A key ingredient of QPE is the 
quantum Fourier transform, which we recall (see e.g. \cite{Mermin:2007}) is defined by
\begin{equation}
	U_{FT}\, |x\rangle = \frac{1}{2^{\frac{n}{2}}}\sum_{y=0}^{2^{n}-1} 
	e^{\frac{2\pi i x y}{2^{n}}} |y\rangle \,, \qquad 0 \le x < 2^{n} \,,
\end{equation}
and which can be shown to be written more explicitly as
\begin{equation}
U_{FT}\, |x\rangle = \frac{1}{2^{\frac{n}{2}}}
\left(|0\rangle + e^{\frac{2\pi i x}{2}}|1\rangle\right)
\otimes \ldots \otimes
\left(|0\rangle + e^{\frac{2\pi i x}{2^{n-1}}}|1\rangle\right)
\otimes 
\left(|0\rangle + e^{\frac{2\pi i x}{2^n}}|1\rangle\right)\,.
\label{QFT}
\end{equation}
The QPE circuit diagram, with $t$ auxiliary qubits, is shown in Fig. \ref{fig:QPE}.

\noindent
At slice 1, one can check with the help of \eqref{QFT} that the circuit 
is in the state 
\begin{equation}
	|\psi\rangle\, U_{FT}\, |\frac{2^{t} \alpha}{2\pi} \rangle \,.
\end{equation}
At slice 2, after applying the inverse quantum Fourier 
transform, the state becomes simply 
\begin{equation}
	|\psi\rangle\, |\frac{2^{t} 
\alpha}{2\pi} \rangle \,.
\label{QPEstate}
\end{equation}
Thus, the measurement of the ancillary qubits 
gives an integer $j$ that is within 1/2 of $\frac{2^{t} \alpha}{2\pi}$.
This estimate can be improved by increasing the value of $t$.

\section{Quantum counting}\label{sec:QC}

We are now in position to address the question raised at the end of 
Sec. \ref{sec:multiplemarked}: what can we do if we are searching for multiple marked elements, but we do not know their number $m$ ?

An answer, known as quantum counting, is to 
apply QPE to the Grover rotation operator; that is,
recalling \eqref{WVmat}, set 
\begin{equation}
    {\mathcal U} = WV = e^{- i 2 \theta\, Y} \,,
    \label{calU}
\end{equation}
and implement the
QPE circuit in Fig. \ref{fig:QPE}
using $|\psi\rangle = |\phi\rangle$ 
\eqref{phi}. Evidently, ${\mathcal U}$ has eigenvalues $e^{\pm 2 i 
\theta}$. Although $|\phi\rangle$ is not an eigenket of ${\mathcal 
U}$, $|\phi\rangle$ lies in the subspace spanned by $|s_{\perp} \rangle$ and $|s 
\rangle$ \eqref{phissperp}; hence, 
QPE provides estimates for both $\pm 2\theta$. More precisely, QPE 
gives integers $j_+$ and $j_-$ that are closest to $\frac{2^t 
(2\theta)}{2\pi}$ and $\frac{2^t (2\pi-2\theta)}{2\pi}$, 
respectively, see \eqref{QPEstate}. 
Solving for $\theta$, we see that
\begin{equation}
	\theta \approx \frac{\pi j_{+}}{2^{t}}\,, \quad \pi - \frac{\pi 
	j_{-}}{2^{t}} \,.
\end{equation}
Using \eqref{sintheta2}, we conclude that the number of marked elements is 
given by
\begin{equation}
m = 2^{n} \sin^2 \theta \approx	2^{n} \sin^2 \left(\frac{\pi 
j_{\pm}}{2^{t}} \right)  \qquad ( \text{ 
implementing } W ) \,.
\label{m1}
\end{equation}

Using $\tilde{W}$ \eqref{Wtilde} instead of $W$ (that is, dropping the sign in \eqref{Wcircuit}) leads to a result different from \eqref{m1}. Indeed, a simple way to account for the different sign is to observe from \eqref{calU}
that changing $W \mapsto -W = \tilde{W}$ implies ${\mathcal U} \mapsto -{\mathcal U}$;
and the corresponding eigenvalues transform accordingly  $e^{\pm 2 i \theta} \mapsto 
-e^{\pm 2 i \theta} = e^{\pm 2 i (\theta \pm \frac{\pi}{2})}$. 
Performing the shift $\theta \mapsto \theta - \frac{\pi}{2}$ in 
\eqref{m1}, we obtain
\begin{equation}
m = 2^{n} \sin^2 \left(\theta -\frac{\pi}{2}\right)
\approx	2^{n} \sin^2 \left[\pi \left(\frac{ 
j_{\pm}}{2^{t}} - \frac{1}{2} \right) \right] \qquad ( \text{ 
implementing } \tilde{W} = -W ) \,.
\label{m2}
\end{equation}

In our simulations, we implemented $\tilde{W}$, and 
the number of marked elements indeed matched with \eqref{m2}.
For example, we searched for the set of {\bf three} 3-bit integers $S=\{ 2, 4, 6\}$ using 5 ancillas, so that $n=m=3$ and $t=5$. In 1024 shots, the two most frequent integers obtained from measuring the ancillas were $j=9$ and $j=23$. Eq. \eqref{m2} then correctly gave $m \approx 3.22 \approx 3$, while \eqref{m1} would imply the incorrect result $m \approx 4.78 \approx 5$.

\section{Conclusion}\label{sec:discuss}

Although quantum counting is not considered part of the standard undergraduate introductory Quantum Computing curriculum (see e.g. \cite{Asfaw:2021vik, Meyer:2023qkd}), it entails only modest -- yet very interesting -- extensions of two topics that {\it are} part
of the standard repertoire, namely Grover's algorithm and QPE. Hence, coding and simulating
quantum counting, using e.g. qiskit \cite{qiskit} or 
cirq \cite{cirq}, is an attractive topic for a student 
project in such a course, especially for students who have already 
coded and simulated Grover and QPE.\footnote{Sample qiskit code is 
included as Supplementary Material; it can be found at {\tt https://arxiv.org/abs/2310.07428}.} 
However, the sign of the diffuser $W$ \eqref{Wcircuit} should not be 
neglected.

\section*{Acknowledgments} 
RN was supported in part by the National Science 
Foundation under Grant No. PHY 2310594 and by a Cooper fellowship.


\providecommand{\href}[2]{#2}\begingroup\raggedright\endgroup

\end{document}